\documentclass[twocolumn]{aastex631}

\usepackage{color, booktabs, subfigure, multirow}
\usepackage{url, hyperref,amsmath}
\usepackage[capitalise]{cleveref}
\usepackage[hang,flushmargin]{footmisc}
\usepackage{xcolor}

\shorttitle{BH mass of 4C+37.11}
\shortauthors{Surti et al.}

\begin{document}

\title{The Central Kinematics and Black Hole Mass of 4C+37.11}

\author{Tirth Surti}
\affiliation{Department of Physics/Kavli Institute for Particle Astrophysics and Cosmology, Stanford University, Stanford, California 94305, USA}

\correspondingauthor{Roger W. Romani}
\email{rwr@astro.stanford.edu}
\author[0000-0001-6711-3286]{Roger W. Romani}
\affiliation{Department of Physics/Kavli Institute for Particle Astrophysics and Cosmology, Stanford University, Stanford, California 94305, USA}

\author[0000-0000-0000-0000]{Julia Scharw\"achter}
\affiliation{Gemini Observatory/NSF's NOIRLab, 670 N’Aohoku Place, Hilo, Hawaii, USA}

\author[0000-0001-8276-0000]{Alison Peck}
\affiliation{Visiting Principal Research Scientist, Department of Astronomy, University of Maryland, 4296 Stadium Dr., College Park, MD 20742-2421}

\author[0000-0001-6495-7731]{Greg B. Taylor}
\affiliation{Department of Physics and Astronomy, University of New Mexico, Albuquerque, NM 87131, USA}


\begin{abstract}
We report on IFU measurements of the host of the radio source 4C+37.11. This massive elliptical contains the only resolved double compact nucleus at pc-scale separation, likely a bound supermassive black hole binary (SMBHB). $i$-band photometry and GMOS-N IFU spectroscopy show that the galaxy has a large $r_b=1.5^{\prime\prime}$ core and that the stellar velocity dispersion increases inside of a radius of influence $r_{\rm SOI} \approx 1.3^{\prime\prime}$. Jeans Anisotropic Modeling analysis of the core infers a total SMBHB mass of $2.8^{+0.8}_{-0.8} \times 10^{10}M_\odot$, making this one of the most massive black hole systems known. Our data indicate that there has been significant scouring of the central kpc of the host galaxy. 
\end{abstract}


\section{Introduction}

\cite{Maness_2004} found that the radio galaxy 4C+37.11 (0402+379) contains two central, compact, flat-spectrum, variable components (designated C1 and C2) with a VLBA-measured separation 7.3 pc at $z=0.055$ and argued that this is a gravitationally bound SuperMassive Black Hole Binary (SMBHB). Subsequent VLBA observations by \citet{2006ApJ...646...49R} have bolstered that claim and even detected a possible relative proper motion of the binary components \citep{2017ApJ...843...14B}. The binary's host is remarkable, being an extremely massive elliptical embedded in a bright X-ray halo \citep{2014ApJ...780..149R, 2016ApJ...826...91A} which represents a cluster's worth of stars and mass in a relatively isolated galaxy. Such `fossil clusters' are believed to be the product of multiple major mergers. We report here on a kinematic study of the central region of this galaxy, which supports a scenario where the source contains a supermassive black hole pair whose present (and former) interactions have scoured the galaxy core. The estimated SMBHB mass is very large, the second largest kinematically measured value in the local universe.

We assume a $H_0 = 69.6$, $\Omega_M = 0.286$, $\Omega_{\Lambda} = 0.714$ cosmology \citep{2014ApJ...794..135B}, so at the redshift of $z=0.055$, 4C+37.11 has a luminosity distance $D_L = 246.9$ Mpc, angular diameter distance $D_A = 221.9$ Mpc, and a projected scale of $1.076$ kpc/arcsec.

\begin{figure*}[]
    \centering
    \includegraphics[scale=0.4]{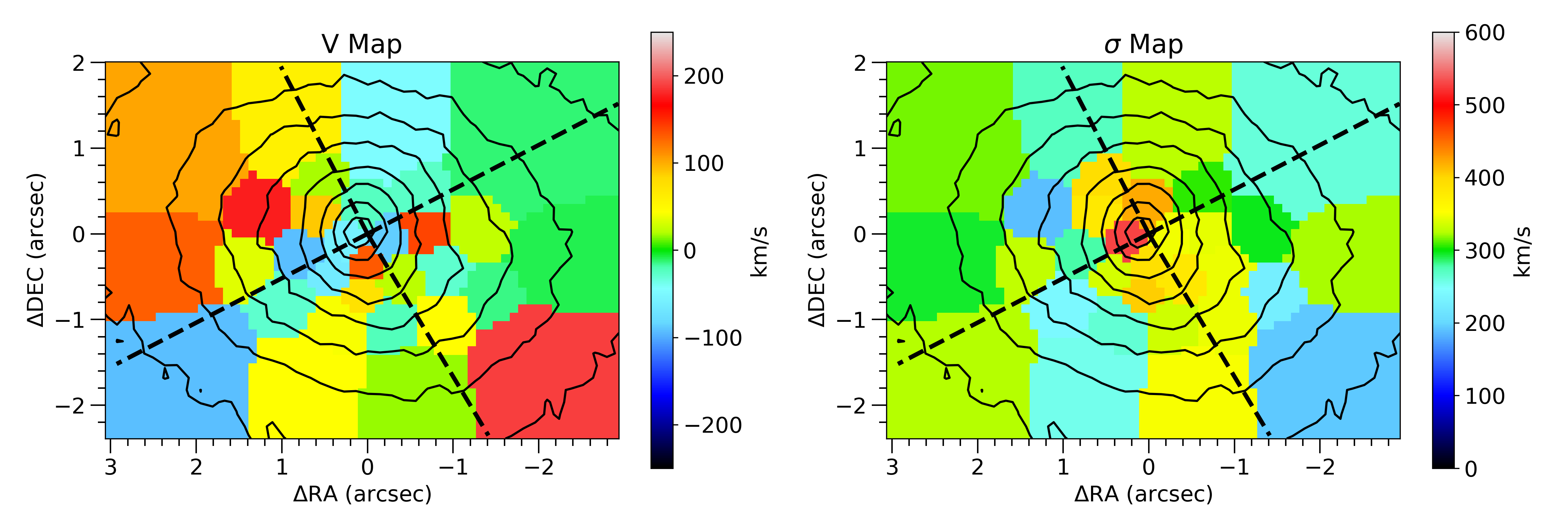}
    \caption{Radial velocity ($V$) and velocity dispersion ($\sigma$) maps of pPXF measurements in the Voronoi bins. The origin is the AGN location, determined from the wings of the [NII] emission lines. Solid contours mark 0.9$\times$,0.8$\times$,...0.2$\times$ the peak surface brightness and dashed lines separate quadrants along the major and minor axes.}
    \label{fig:kin_maps}
\end{figure*}

\begin{figure}[h!]
    \begin{subfigure}
    \centering
    \includegraphics[scale=0.45]{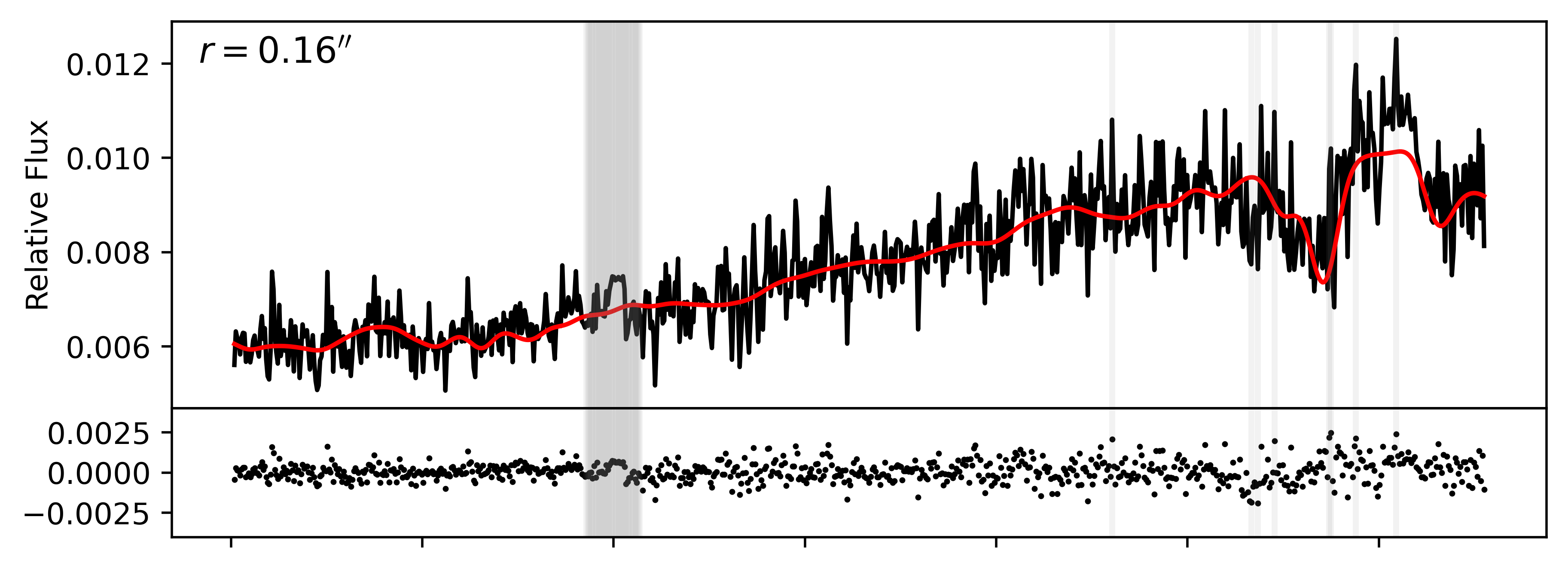}
    \end{subfigure}
    
    \vspace{-3mm}
    \begin{subfigure}
    \centering
    \includegraphics[scale=.45]{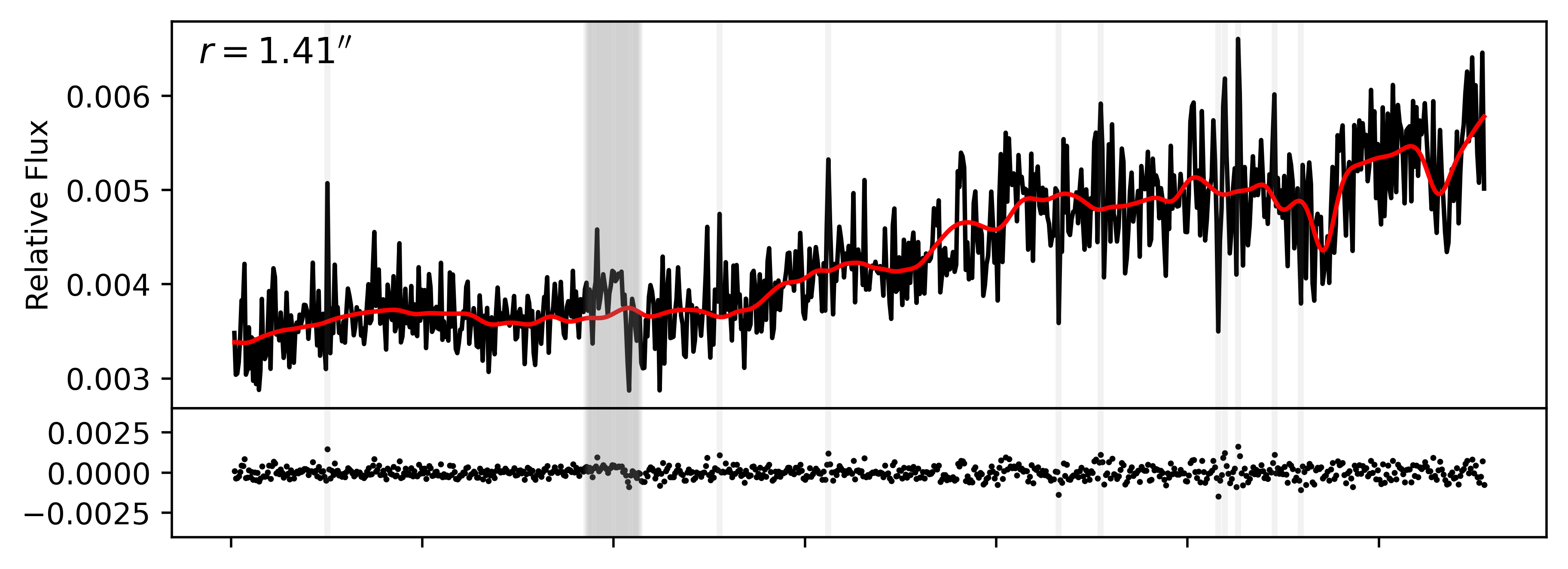}
    \end{subfigure}

    \vspace{-3mm}
    \begin{subfigure}
    \centering
    \includegraphics[scale=.45]{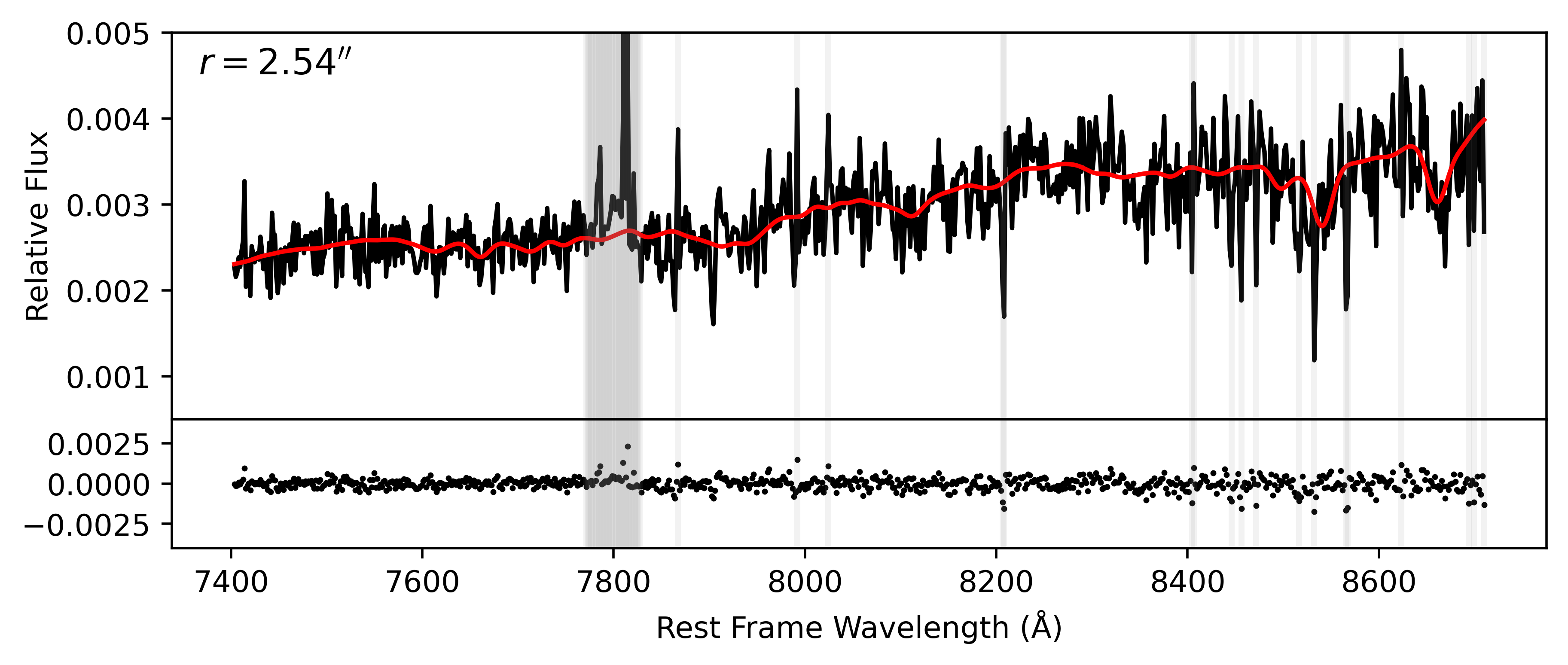}
    \end{subfigure}
    \caption{Sample spectral fits (red) and residuals for S/N $\sim 20$ Voronoi bins at three different radii. For plotting purposes, the spectra were rebinned to $61.2$ km/s, 4$\times$ the width used for the kinematic extraction. Grey bars show masked pixels from chip gaps and from 2-$\sigma$ iterative clipping.}\label{fig:spectral_fits}
\end{figure}

\section{Observations, Data Reduction and Measurement}

Our kinematic study uses two observations. A 2008 December 12 WIYN 3.6m 2200\,s OPTIC $i$ band image presented in \cite{2014ApJ...780..149R} has image quality FWHM $<0.6^{\prime\prime}$ and measures the galaxy surface brightness profile. For spectroscopy we use a 2015 December 1 Gemini 8.1\,m $6\times 1200$\,s GMOS-N IFU observation of the galaxy center, downloaded from the Gemini Science archive. These observations used the B600 grating, covering 6300--9195\AA\,at dispersion 0.47 \AA/pixel. Conditions were good, with low airmass throughout. The IFU feeds a fiber array with a $0.22^{\prime\prime}$ pitch, and with the observations dithered on the sky, IRAF GMOS pipeline processing calibrated and cleaned the spectra, assembling a data cube covering $4.5^{\prime\prime} \times 6.2^{\prime\prime}$ at $0.1^{\prime\prime}$ scale, centered on the AGN, with reduced net exposure at the edges. Measurements of images of the PSF-distributed line wings of the AGN core find an angular resolution of $0.64^{\prime\prime}$ (FWHM) and sky lines give a median effective spectral resolution of $2.0$ \AA\,FWHM.

For our kinematic study, we fit the emission line-free range 7810--9190\AA\,, masking a small range near 8230\AA\ affected by a chip gap. We log rebin the spectra over this range to a velocity scale of $15.3$ km/s. Reliable measurements of the line-of-sight velocity distribution (LOSVD) require a signal-to-noise (S/N) of $\sim 20$/resolution element. Our data cube provides only S/N $\sim$ 7/spaxel in the galaxy center, falling to S/N $\sim 1$ in the more poorly exposed outskirts of the data cube. Thus we use Voronoi binning \citep{2003MNRAS.342..345C} to combine the spaxels. With a target S/N $\sim 20$ at the rebinned velocity scale we get 31 bins (Fig.\,\ref{fig:kin_maps}); edge bins generally do not reach the S/N target. Imperfectly excised cosmic ray events and poor sky subtraction affect many of the edge pixels with few IFU exposures. We therefore used iterative sigma clipping to excise outlier pixels from each wavelength in the Voronoi tiles. The means of the clipped tile wavelength measurements form the final spectra, with their dispersion as the associated error vector. Velocity structure is measured with the Penalized PiXel Fitting method (pPXF) of \cite{2023MNRAS.526.3273C}, which convolves template stellar spectra with a Gauss-Hermite decomposition of the LOSVD. With the limited S/N we fit only $V$ and $\sigma$, skipping the higher order $h_3$ and $h_4$ moments.

For the stellar templates, we used a combination of G0--M2 giants from the Indo-US Stellar Library \citep{Valdes_2004}. These templates have a slightly finer scale of 0.44 \AA/pixel, with a median spectral resolution of $\sim 1$ \AA. To match the templates to the host galaxy spectra, we convolve the templates using the difference in the spectral resolutions, following Section 2.2 in \citep{10.1093/mnras/stw3020}. pPXF uses additive and multiplicative polynomials in the fitting model for both the target and template spectra to reduce sensitivity to imperfect continuum fluxing and veiling components. We varied the order of these polynomials, finding that both additive and multiplicative terms required order $\geq 4$ to achieve stable fits. Here we use 4$^{th}$-order additive and 6$^{th}$-order multiplicative Legendre polynomials (4A+6M) as fiducial. Sample spectral fits at three different radii are shown in Figure \ref{fig:spectral_fits}.

Radial velocity and velocity dispersion maps from the Voronoi tiles are shown in Figure \ref{fig:kin_maps}. 
We detect little if any coherent rotation, typical for massive ellipticals. However, given the large size of the Voronoi bins, we cannot exclude a kinematically distinct core, sometimes seen among non-regular rotators \citep{Cappellari_2016}. We do see a substantial Keplerian-like peak in $\sigma(r)$ which flattens out at $\sim$300\,km/s beyond $\sim 2^{\prime\prime}$. Although we lack kinematic data at the half-light radius ($R_e \sim 12^{\prime\prime}$, see below in the discussion of the Multi-Gaussian Expansion), we can follow \citet[][App.\,A]{2016ApJ...818...47S} and estimate the effective velocity dispersion by integrating the surface brightness-weighted $V_{rms}$ to our outermost bin, finding $\sigma_{e} = 332 \pm 11$\,km/s.  This likely overestimates $\sigma_e$, as over a third of our bins are affected by the central velocity cusp.

\begin{figure}[h!]
    \begin{subfigure}
    \centering
    \includegraphics[scale=0.45]{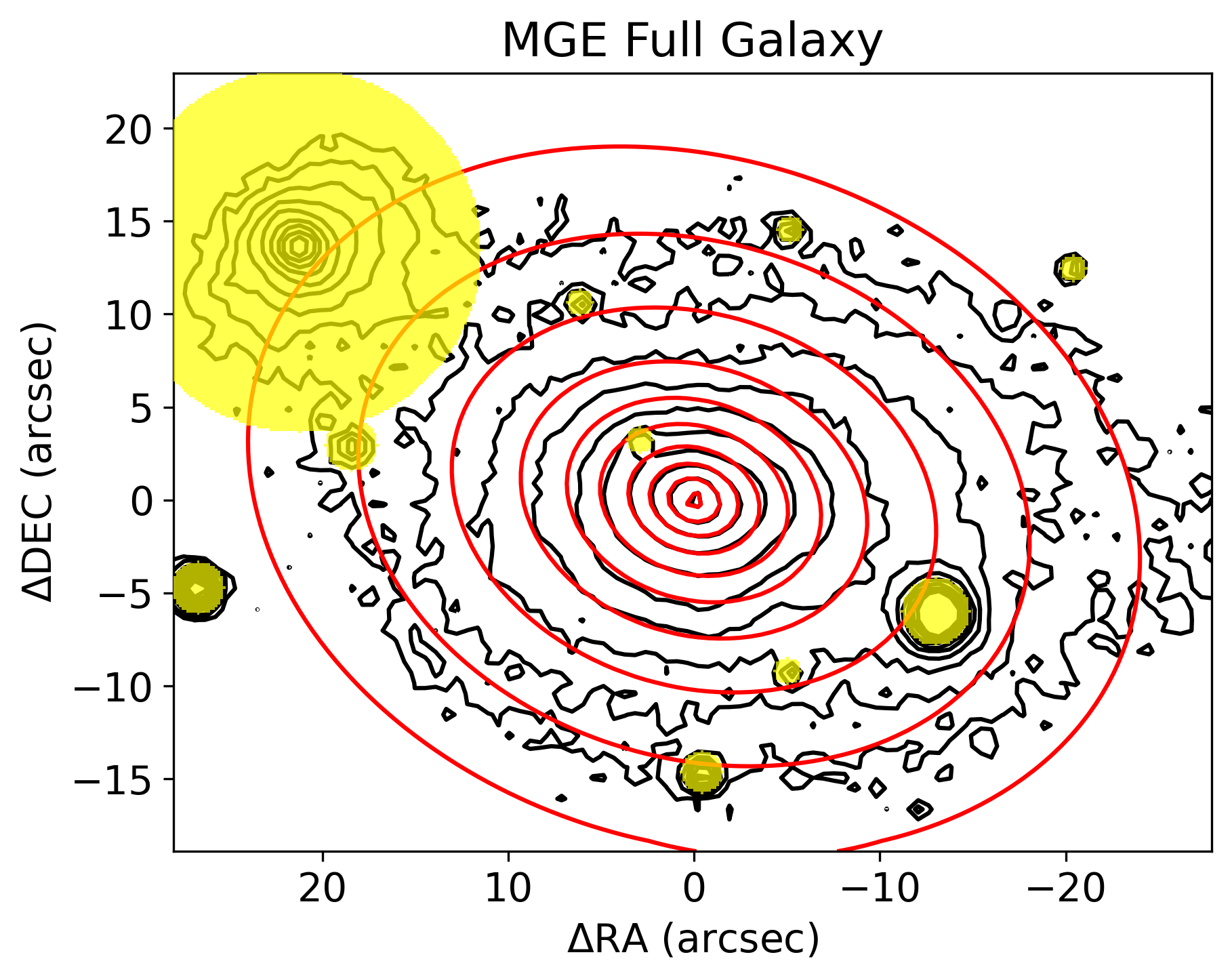}
    \end{subfigure}

    \vspace{-3mm}
    \begin{subfigure}
    \centering
    \includegraphics[scale=.45]{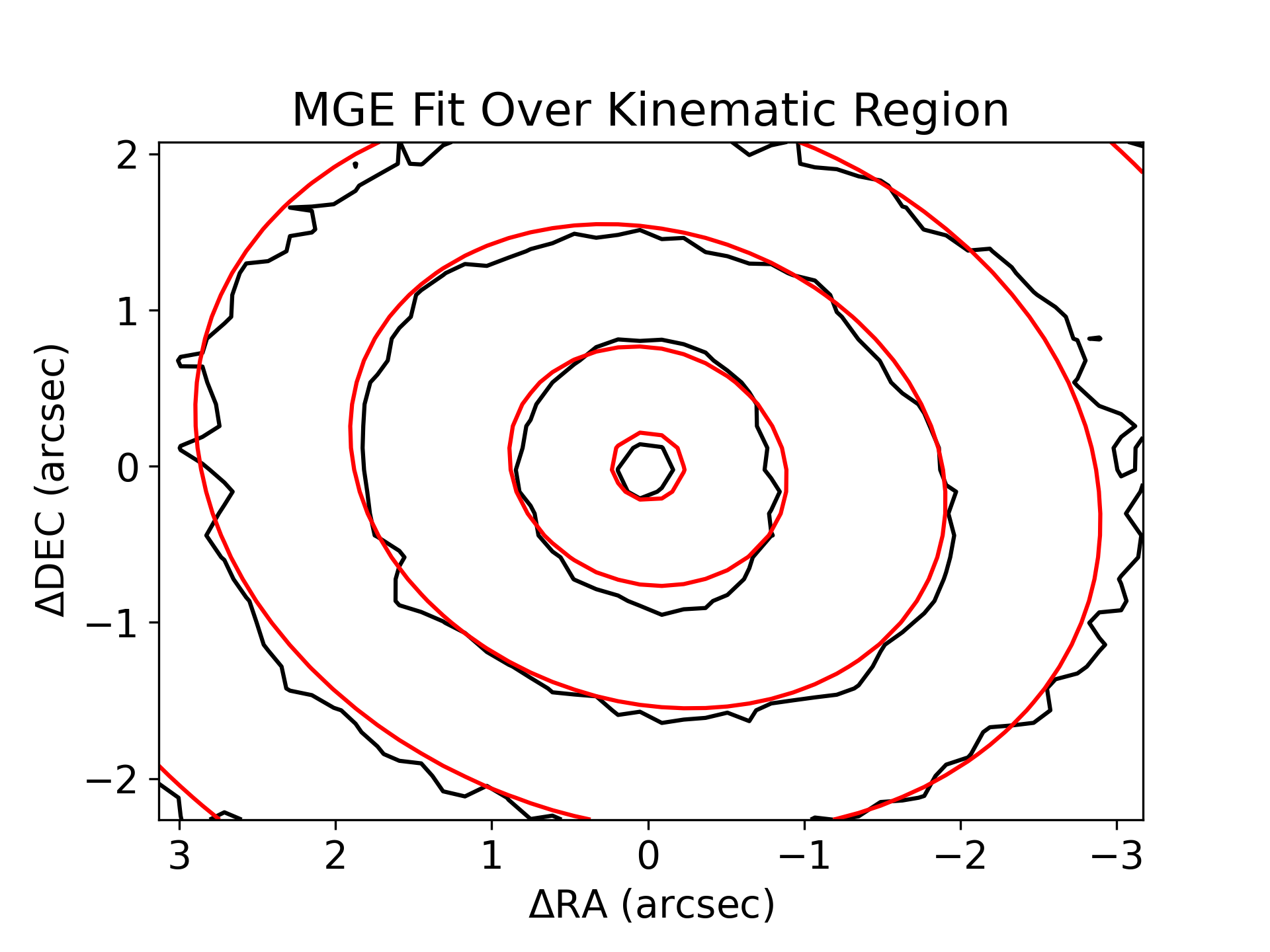}
    \end{subfigure}
    
    \caption{$i'$ image isophote contours (black) and MGE fit ellipses (red). Top: the full region, Bottom: the IFU region with kinematic measurements. The yellow regions correspond to masked pixels in the MGE fits.}\label{fig:example}   
\end{figure}

For dynamical modeling, we also require the host surface brightness profile. This was measured from the WIYN $i'$ image using the Multi-Gaussian Expansion (MGE) method of \cite{Cappellari2002}. We first masked stars and galaxies near 4C+37.11. We then obtained the image PSF from 15 unsaturated stars near the host, using the \texttt{photutils} effective PSF (ePSF) routine \citep{larry_bradley_2023_7946442}. In host fitting we assume axisymmetry, with the surface brightness modeled as a sum of elliptical Gaussian components aligned along the direction of the photometric major axis ($x'$):
\begin{equation}\label{eq1}
    \Sigma (x', y') = \sum_{j=1}^{N}S_j {\rm exp}\left[-\frac{1}{2\sigma_i}\left(x'^2 + \frac{y'^2}{q_i^2}\right)\right].
\end{equation}
Here we have $N$ Gaussian components with surface brightness $S_j$ and axial ratio $q_j$. The position angles of the Gaussians were fixed at $PA_{\rm MGE}=72.4^\circ$ and $q_j$ had a flat prior 0.75--1. Before fitting we convolve \eqref{eq1} with a MGE expansion of the WIYN PSF. The MGE fits for the PSF and the i' band image are summarized in Tables \ref{mgefit_1} and \ref{mgefit_2}, respectively. Figure \ref{fig:example} shows the MGE fits for the IFU region and the entire region used in mass modeling. While somewhat triaxial at large radii, the axisymmetric assumption is adequate over the IFU kinematics region. Using the MGE, with the routine \texttt{mge\_half\_light\_isophote} of the package \texttt{JamPy} of \citep{Cappellari2008} we find a half-light radius for 4C+37.11 to be $R_e = 12.1^{\prime\prime}.$

\begin{table}[]
\centering
\begin{tabular}{cccc}
\hline
$j$ & $S_j$ & $\sigma_j$  ($\mathrm{arcsec}$) & $q_j$ \\ \hline
1   & 0.23  & 0.15                            & 0.71  \\
2   & 1.38  & 0.27                            & 0.93  \\
3   & 0.362 & 0.36                            & 1.00  \\
4   & 0.10  & 0.56                            & 0.96  \\
5   & 0.02  & 0.94                            & 1.00  \\
6   & 0.002 & 2.1                             & 1.00 
\end{tabular}
\caption{WIYN PSF MGE} \label{mgefit_1}
\end{table}

\begin{table}[]
\centering
\begin{tabular}{cccc}
\hline
$j$ & $\log(S_j)$  ($L_{\odot}$ $\mathrm{pc}^{-2}$) & $\sigma_j$  ($\mathrm{arcsec}$) & $q_j$ \\ \hline
1   & 4.18                                          & 0.11                            & 0.75  \\
2   & 3.18                                          & 1.25                            & 0.75  \\
3   & 2.83                                          & 3.02                            & 0.76  \\
4   & 2.29                                          & 6.38                            & 0.75  \\
5   & 2.01                                          & 17.8                            & 0.75 
\end{tabular}
\caption{PSF-Convolved 4C+37.11 MGE}\label{mgefit_2}
\end{table}

It was noted in \citet{2014ApJ...780..149R} that the host surface brightness profile flattened within $\sim1.5$ kpc from an outer $n=4$ S\'ersic profile, indicting that 4C+37.11 is a cored elliptical. We measure the core radius $r_b$ by fitting a core-S\'ersic model, with a point source excess for the AGN light, to the $i'$ band image, using the reconstructed PSF for model convolution and minimizing with the 2D fitting software \texttt{imfit} \citep{Erwin_2015}. The core-S\'ersic profile is described as 
\begin{equation}
    I(r) = I'\left[1 + \left(\frac{r_b}{r}\right)^\alpha\right]^{\frac{\gamma}{\alpha}}\mathrm{exp}\left[-b_n\left(\frac{r^\alpha + r_b^{\alpha}}{r_e^{\alpha}}\right)^{\frac{1}{\alpha n}}\right]
\end{equation}
where
\begin{equation}
    I' = I_{b}2^{-\gamma/\alpha}\mathrm{exp}\left[b_n\left(2^{1/\alpha}r_b/r_e\right)^{1/n}\right]
\end{equation}
as per \cite{Graham_2003}. For our fit we fix the sharpness parameter at $\alpha=100$ \citep{Graham_2003,2004AJ....127.1917T} and derive the fit parameters in Table \ref{tab:surf_bright_fits}, where we have converted intensity at the break radius to a surface brightness $S_{b,i'}$. We do indeed see that the outer galaxy has $n\approx 4$ and find a core radius $r_b = 1.60 \pm 0.05$\,kpc. For $\alpha>10$, we find that $r_b$ varies by $<2\sigma$ from this value.

\begin{table}[h!]
\centering
\begin{tabular}{ll}
\hline
Core S\'ersic Parameters & Value             \\ \hline
PA (deg)                                & $81.9 \pm 0.3$    \\
$\epsilon$                             & $0.293 \pm 0.003$ \\
$S_{b, i'}$ (mag $\mathrm{arcsec}^{-2}$)       & $18.26 \pm 0.03$  \\
$n$                                     & $3.92 \pm 0.08$   \\
$r_e$ (arcsec)                          & $19.0 \pm 0.5$    \\
$r_b$ (arcsec)                          & $1.49 \pm 0.05$   \\
$\gamma$                                & $0.40 \pm 0.03$   \\ \hline
Point Source Parameters                 & Value             \\ \hline
$i'_{tot}$ (mag)                         & $20.2 \pm 0.3$   
\end{tabular}
\caption{2D Core-S\'ersic + Point Source model fit parameters. Magnitudes $S_{b, i'}$ and $i'$ are corrected for an estimated $A_{i'}=2.2$\,mag extinction.}
\label{tab:surf_bright_fits}
\end{table}

\section{Dynamical Modeling}

Given the low S/N kinematics data extending only to $\sim 3^{\prime\prime}$, we do not attempt full Schwarzschild orbit-superposition modeling. Instead, we conducted axisymmetric Jeans Anisotropic Modeling (JAM) using the software \texttt{JamPy} \citep{Cappellari2008}, which does not require large-radii kinematics to achieve good constraints \citep{2024MNRAS.527.2341S} and produces BH masses consistent with Schwarzschild analyses \citep{Cappellari_2010}.

Unlike the Schwarzschild method, axisymmetric JAM makes a number of assumptions about the distribution of stellar velocities. In particular, the velocity ellipsoid is either assumed to be aligned with spherical $(r, \phi, \theta)$ or cylindrical $(R, \phi, z)$ coordinates, neither of which can fully represent a real galaxy. Spherically-aligned velocity ellipsoids are typically good approximations to slow rotator ellipticals, which are weakly triaxial and have spherical isophotes in the inner half-light radius \citep{2024MNRAS.527.2341S}. Cylindrically-aligned velocity ellipsoids are typically applied to fast rotator massive ellipticals \citep{Cappellari2008} but have also been assumed for slow and non-regular rotators, as in the ATLAS3d survey \citep{Cappellari_2013}.
Here, we compare JAM models using both a cylindrically-aligned and spherically-aligned velocity ellipsoid.
In the cylindrical case, we assume radial anisotropy $\beta_z = 1 - \sigma_z^2 / \sigma_R^2$ is constant with $\sigma_R^2$ = $b \sigma_z^2$. In the spherical case, radial anisotropy $\beta = 1 - \sigma_{\theta}^2/\sigma_{r}^2$ is constant. The JAM models include no tangential anisotropy $\gamma = 1-\sigma_{\phi}^2 / \sigma_R^2 = 0$, since the velocity second moment predictions are independent of $\gamma$; nonzero tangential anisotropy is not usually needed to model the overall kinematics of real galaxies \citep{Cappellari_2016}. 

\subsection{JAM Models}

For both the spherically and cylindrically aligned cases, we incorporate a \textbf{Self-Consistent Model}, assuming that the total mass density follows the surface brightness profile. In particular, the projected mass density can be obtained by multiplying each Gaussian component in \eqref{eq1} by a constant mass-to-light ratio $\Upsilon_{i'}$. The parameters for this model are then $\Upsilon_{i'}$, the radial anisotropy parameter ($\sigma_z / \sigma_R$ for cylindrical or $\sigma_{\theta} / \sigma_{r}$ for spherical), inclination $i$, and the black hole mass $M_{\bullet}$. Here inclination $i$ is the angle between the galaxy's symmetry axis and the line-of-sight direction and $M_{\bullet}$ represents the total mass of the binary which is modeled by a Keplerian point mass potential. We represent the PSF of the kinematic observations with a single circular Gaussian of dispersion $0.27^{\prime\prime}$ (as described in Section 2). We do not incorporate a model for the contribution of the dark halo's potential due to the limited S/N and radial range of the kinematic data; tests with a NFW halo give no halo constraints and negligible changes to the other fit parameters.

At S/N $\sim 20$ the central bin has a large $V_{rms} =\sqrt{V^2 + \sigma^2}\approx 532$ km/s (Fig.\,\ref{fig:vrms_model}). Since this has a strong effect on the black hole mass fit, we tested for sensitivity and systematic bias by optionally dropping the innermost bin. We also fit using a lower Voronoi target S/N $=18$, for additional bins across the central $0.5^{\prime\prime}$. In all cases the $V_{rms}$ rise from the SMBHB is still apparent.

\begin{table*}[t]
\centering
\begin{tabular}{cccccccc}
\multicolumn{1}{l}{S/N} & \multicolumn{1}{l}{Velocity Ellipsoid} & \multicolumn{1}{l}{\begin{tabular}[c]{@{}l@{}}Central\\ Kin. Bin\end{tabular}} & \multicolumn{1}{l}{$i \text{ (deg)}$} & \multicolumn{1}{l}{$\sigma_{\theta}/\sigma_{r}$} & \multicolumn{1}{l}{$\sigma_{z}/\sigma_{R}$} & \multicolumn{1}{l}{$M_{\bullet}$ ($M_{\odot}$)} & \multicolumn{1}{l}{$\Upsilon_{i'}$ $(M_{\odot} / L_{\odot, i'})$} \\ \hline
20                      & Spherical                              & Kept                                                                               & $73^{+12}_{-14}$                     & $1.0^{+0.3}_{-0.2}$                              & -                                           & $(2.8^{+0.8}_{-0.8})\times 10^{10}$             & $4.9^{+0.9}_{-0.7}$                                              \\
20                      & Spherical                              & Removed                                                                            & $71^{+13}_{-15}$                     & $0.83^{+0.23}_{-0.18}$                           & -                                           & $(2.0^{+0.9}_{-0.9})\times 10^{10}$             & $4.7^{+0.8}_{-0.7}$                                              \\
20                      & Cylindrical                            & Kept                                                                               & $66^{+15}_{-13}$                     & -                                                & $0.92^{+0.09}_{-0.10}$                      & $(2.7^{+0.5}_{-0.4})\times 10^{10}$             & $4.8^{+0.5}_{-0.5}$                                              \\
20                      & Cylindrical                            & Removed                                                                            & $66^{+17}_{-15}$                     & -                                                & $0.91^{+0.08}_{-0.10}$                       & $(2.4^{+0.5}_{-0.5})\times 10^{10}$             & $4.9^{+0.5}_{-0.5}$                                              \\
18                      & Spherical                              & Kept                                                                               & $72^{+12}_{-14}$                     & $0.85^{+0.20}_{-0.16}$                           & -                                           & $(2.0^{+0.6}_{-0.6}) \times 10^{10}$            & $4.1^{+0.7}_{-0.5}$                                              \\
18                      & Spherical                              & Removed                                                                            & $72^{+12}_{-15}$                     & $0.83^{+0.18}_{-0.16}$                           & -                                           & $(1.9^{+0.6}_{-0.7}) \times 10^{10}$            & $4.1^{+0.6}_{-0.5}$                                              \\
18                      & Cylindrical                            & Kept                                                                               & $66^{+15}_{-15}$                     & -                                                & $0.89^{+0.08}_{-0.10}$                      & $(2.3^{+0.3}_{-0.3})\times 10^{10}$             & $4.2^{+0.4}_{-0.4}$                                              \\
18                      & Cylindrical                            & Removed                                                                            & $68^{+15}_{-16}$                     & -                                                & $0.89^{+0.07}_{-0.09}$                      & $(2.2^{+0.4}_{-0.4})\times 10^{10}$             & $4.2^{+0.4}_{-0.4}$                                             
\end{tabular}
\caption{The Self-Consistent Model posterior medians and 1-$\sigma$ ranges for the different kinematic models/data sets.}
\label{table:post_modes}
\end{table*}

\subsection{Results}
We perform MCMC fitting of the JAM models,  with a $sin(i)$ prior on the inclination and wide uniform priors on the other parameters. The marginal posterior values and 68\% confidence regions are summarized in Table \ref{table:post_modes} for two S/N binnings, both velocity ellipsoid models, and retention or exclusion of the central kinematic point.
We note that while error bars are larger on the spherical model parameters and while low S/N binning tends to slightly reduce the best-fit black hole masses, all results are consistent at the $\sim 1\sigma$ level.

\begin{figure}[h!]
    \centering
    \includegraphics[scale=0.4]{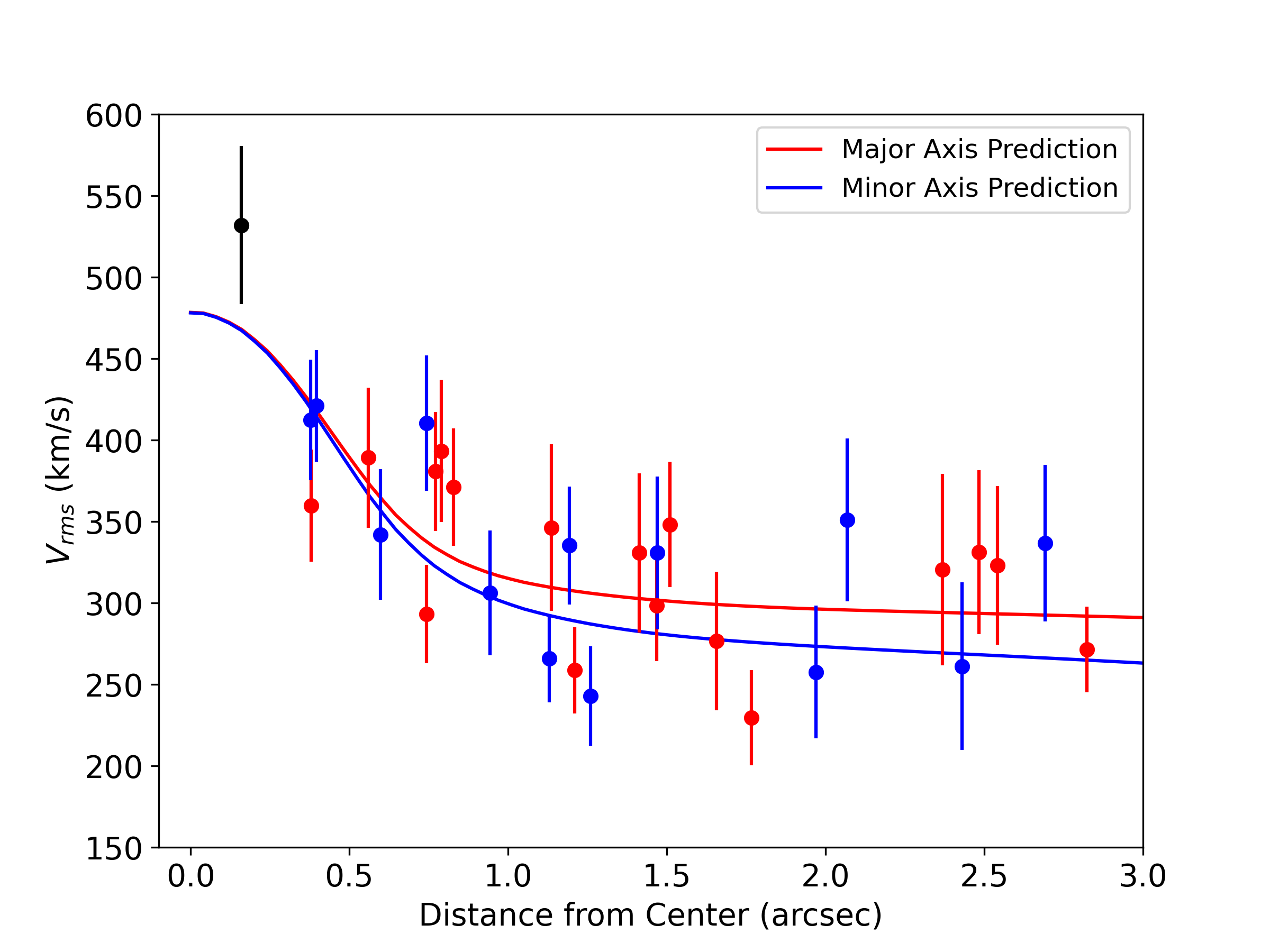}
    \caption{$V_{\rm rms}=\sqrt{V^2 + \sigma^2}$ $vs.$\, radius compared with the best-fit spherical JAM model. While the fit was to all data, the points are color-coded by major axis/minor axis sector (Fig.\,1), and the curves show the expected model runs along the major and minor axes. The $V$ and $\sigma$ values are available as a DbF file.}
    \label{fig:vrms_model}
\end{figure}

Some additional systematic sensitivity comes from the choice of the spectrum/template polynomials. We explored this by re-fitting the cylindrical $\text{S/N}=20$ model (with its smaller statistical errors), varying the degree of multiplicative and additive polynomials from 4 through 7 (16 combinations). Since the SMBHB mass $M_\bullet$ is our prime measured parameter, we focus on its sensitivity. While over half of the values fell within $0.5\times 10^{10}M_\odot$ of the 4A+6M fit result, values ranged from $2.25-3.75\times 10^{10} M_\odot$. Evidently, significant systematic uncertainties remain in the velocity extractions; these can be controlled with higher S/N data. 

\begin{figure}[h!]
    \centering
    \includegraphics[scale=0.35]{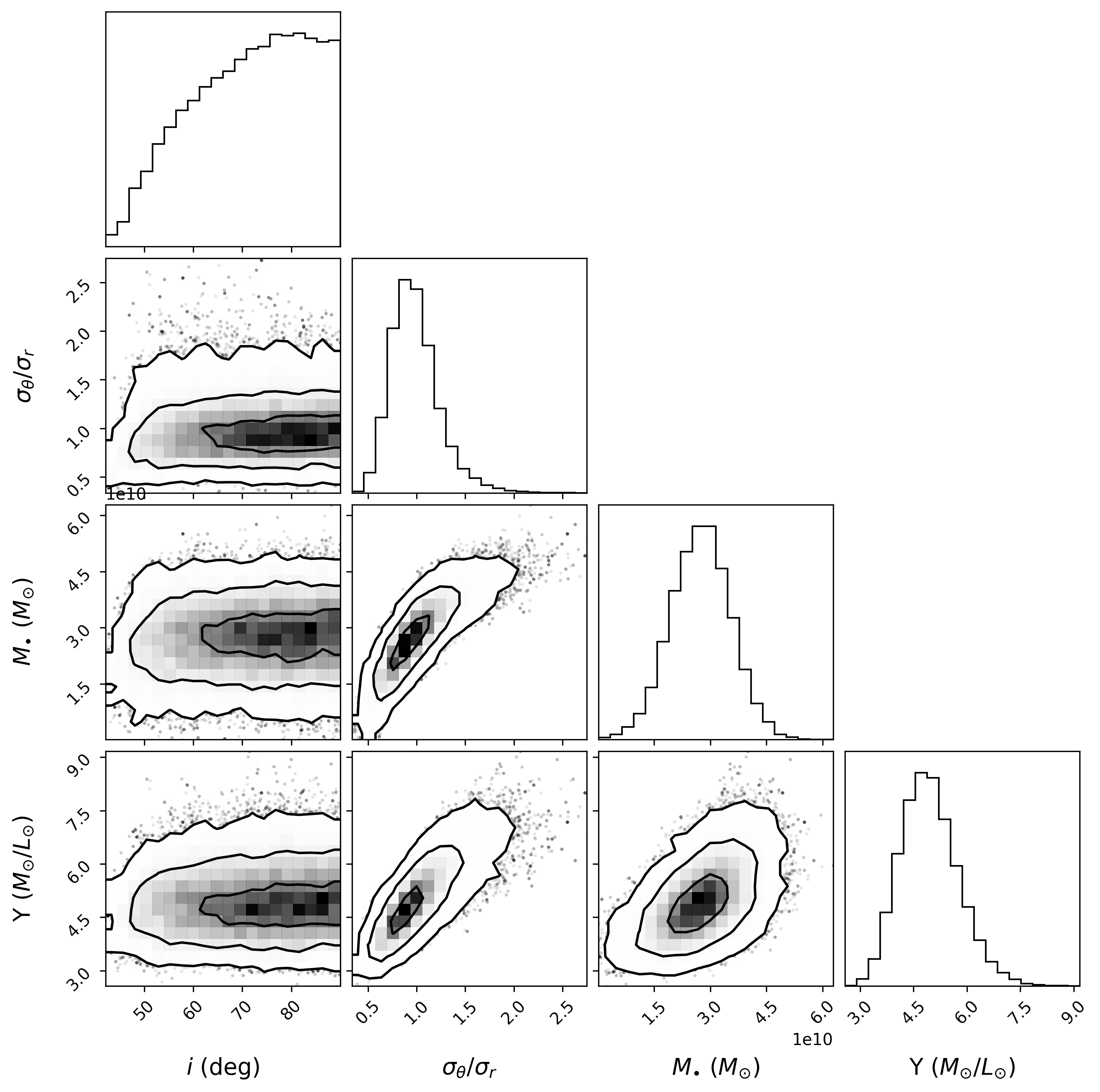}
    \caption{Corner plot for the Spherical S/N=20 JAM fit.}
    \label{fig:corner}
\end{figure}

Since 4C+37.11 appears to be a slow rotator, we take the spherical S/N=20 fits as our standard, giving $M_{\bullet} = (2.8\pm0.8) \times 10^{10}$ $M_{\odot}$, statistical errors only. While the mass is large, so are the error bars, encompassing nearly all of the other fits and polynomial choices. However, a slightly more conservative interpretation might adopt $M_\bullet \approx 2.4 \times 10^{10}M_\odot$, about $\sigma/2$ lower, since the central S/N=20 bin does tend to increase the mass. The radial $V_{\rm rms}$ run and corner plots for the spherical fit are shown in Figs.\,\ref{fig:vrms_model} and \ref{fig:corner}. In the latter we note that $i$ is only weakly constrained within the prior and that $M_\bullet$ is correlated with the velocity anisotropy, with larger hole masses demanded for $\beta<0$. Higher S/N is clearly needed to pin down $\beta$ (and its possible radial dependence due to core scouring, see below).

\begin{figure*}[t!!]
    \centering
    \subfigure[]{\includegraphics[width=0.49\textwidth]{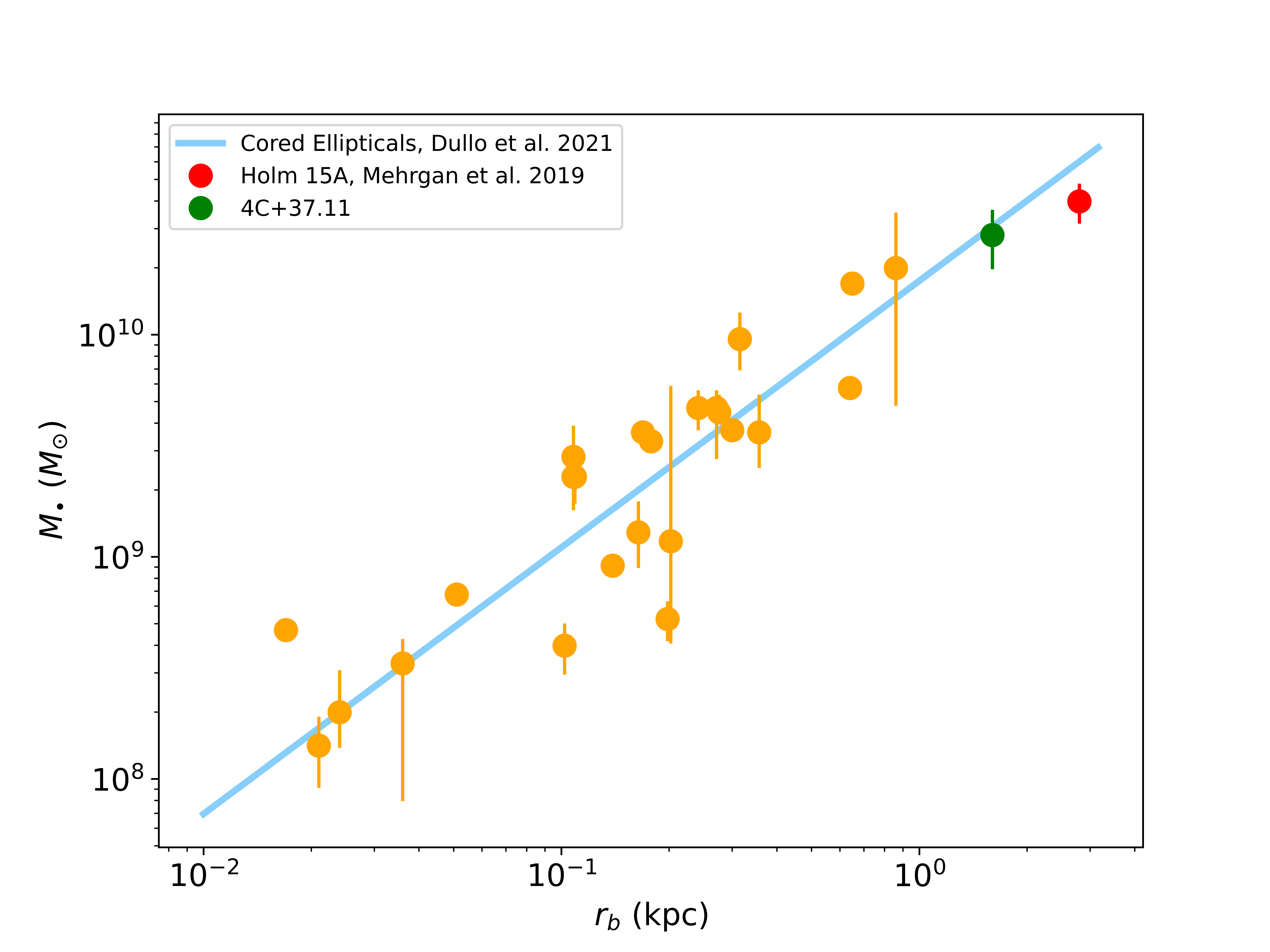}}
    \subfigure[]{\includegraphics[width=0.49\textwidth]{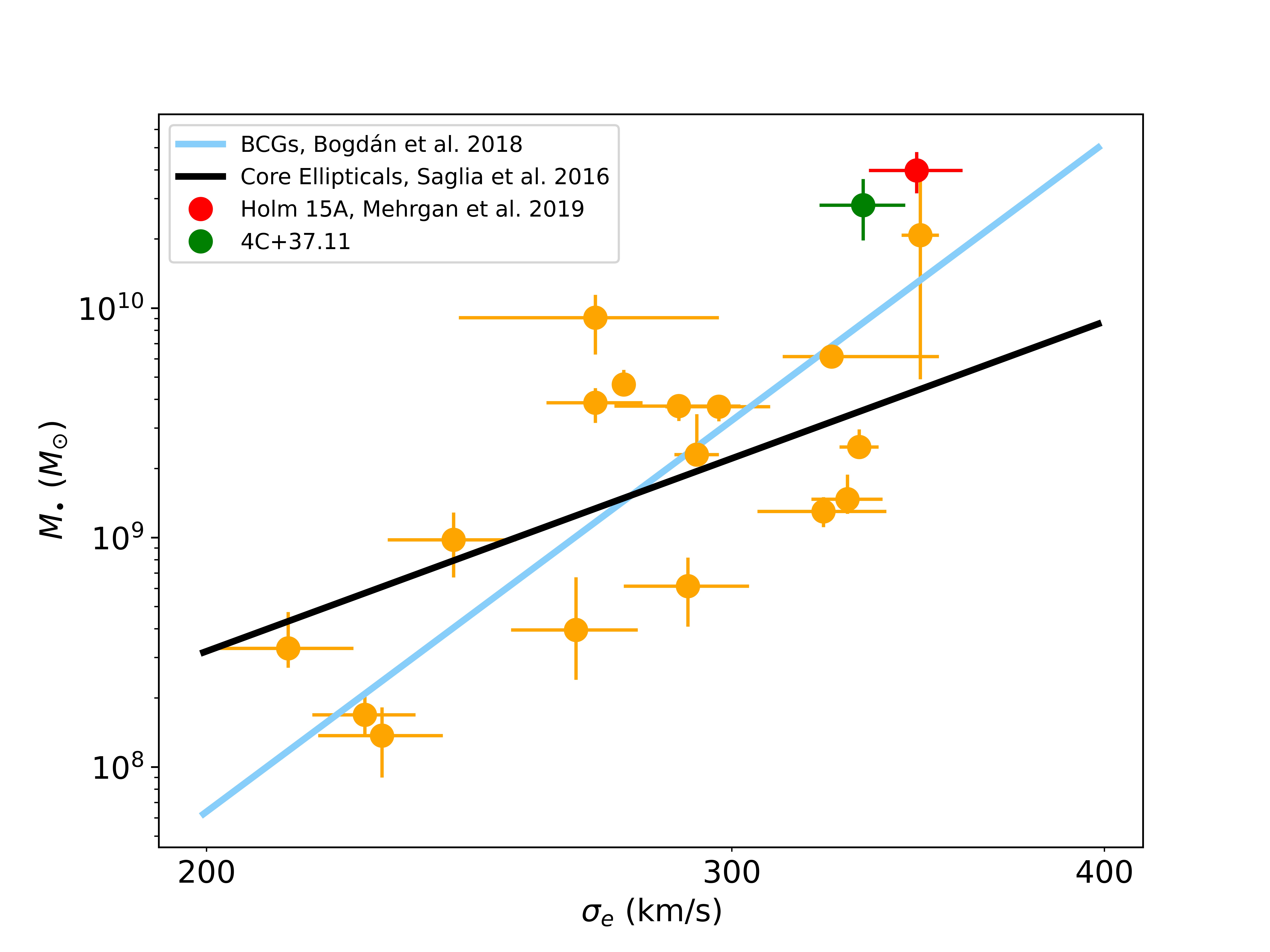}} 

\caption{(a) Correlation between core radius and black hole mass from \cite{2021ApJ...908..134D}, including measurement of depleted core Holm 15A from \cite{Mehrgan_2019}. (b) Correlation with effective velocity dispersion for BCGs from \cite{2018ApJ...852..131B}, represented by the orange data points and blue fit line. The black line is the trend for all core ellipticals from \cite{2016ApJ...818...47S}. Note that 4C+37.11/Holm 15 lie well above the relation for all ellipticals, but are consistent with the BCG trend.
}
    \label{fig:foobar}
\end{figure*}

Using the black hole mass, constant stellar mass-to-light ratio, and inclination for the spherical model, we estimate the radius of the sphere of influence $r_{SOI}$ of the binary. Following \cite{Mehrgan_2019} and \cite{2024MNRAS.527.2341S}, we define $r_{SOI}$ by $M_{\star}(\leq r_{SOI}) = M_{\bullet}$, where $M_{\star}$ is the enclosed stellar mass. Using the routine \texttt{mge\_radial\_mass} of \texttt{JamPy} \citep{Cappellari2008} to compute the enclosed stellar mass within a radius $r$ using the MGE of the surface brightness, we find $r_{SOI} \approx 1.29^{\prime\prime}$ ($1.38$\,kpc), quite consistent with the observed $V_{rms}$ flattening.

\section{Discussion and Conclusions}

It is interesting to compare our dynamical $M_\bullet \approx 2.8\times 10^{10} M_\odot$ estimate for 4C+37.11's binary with values found for other massive ellipticals. These $M_\bullet$ values correlate with host properties, especially those of the host core, giving insight into black holes/host co-evolution \citep{Kormendy_2013,2016ApJ...818...47S,2018ApJ...852..131B}. 

The core radius $r_b$ has been found to tightly correlate to the central black hole mass among massive ellipticals \citep{Mehrgan_2019,2021ApJ...908..134D}. We compare our 4C+37.11 values with dynamically measured $M_\bullet$ and $r_b$ for other elliptical galaxies \citep{2021ApJ...908..134D}, including Holm 15A, which has the largest dynamically measured black hole mass in the local universe, from \cite{Mehrgan_2019}. We find our value is quite consistent with these authors' best-fit trend for cored ellipticals (Figure \ref{fig:foobar}a).

Another well-known correlation is the $M_\bullet-\sigma$ relation, which has been found to depend on the host type. While our  $\sigma_e = 332\pm 11$\,km/s is somewhat suspect since we can only integrate to $\approx 0.25R_e$, we can compare with the trends seen by various authors. We first consider the cored ellipticals with dynamical mass measurements in \cite{2016ApJ...818...47S}. Our estimate exceeds their best-fit correlation's prediction by $\sim 7.2\times$. However, if we compare with the cored ellipticals which are also brightest cluster galaxies (BCGs) from \cite{2018ApJ...852..131B}, we find that 4C+37.11 (and Holm 15A) follow their best-fit trend more closely (Figure \ref{fig:foobar}b).

\bigskip
The $M_\bullet \approx 2.8\times 10^{10} M_\odot$ dynamical mass for 4C+37.11's SMBHB is one of the largest measured in the local universe. While exceeding the value predicted for a typical elliptical, it is in line with expectations for cored ellipticals and, especially, those that are also BCGs. This fits well with the picture that the host is a fossil cluster, the product of several major mergers. If these mergers were largely dry, dissipation-less events, they would grow the central black hole mass faster than the stellar velocity dispersion \citep{Mehrgan_2019}. In addition, the back-action from the current binary (and likely from earlier binary phases) has `scoured' the core, removing stars capable of exerting dynamical friction via three-body gravitational slingshots \citep{1980Natur.287..307B}. In that, 4C+37.11 resembles other extreme cored ellipticals, including Holm 15A. This picture can be tested and extended by additional kinematic studies, which can measure $\beta(r)$, tighten the $M_\bullet$ estimate and constrain the orbit anisotropy of the stars in the host core, probing the system's merger history. While such measurements will be challenging or impossible from the ground, the {\it JWST} NIRSpec IFU field is well matched to the host $r_b$ and can provide important information on the core kinematics.
\bigskip

We thank the anonymous referee for constructive comments which substantially improved the paper and thank the Stanford on the Moon program for publication support. AP acknowledges the National Science Foundation for providing support to staff for independent research.
This paper used data from program GN-2015B-FT-6, which was recovered from the Gemini Observatory Archive and processed with the Gemini IRAF package. Thus this work is based in part on observations obtained at the international Gemini Observatory, a program of NSF's NOIRLab, which is managed by the Association of Universities for Research in Astronomy (AURA) under a cooperative agreement with the National Science Foundation on behalf of the Gemini Observatory partnership: the National Science Foundation (United States), National Research Council (Canada), Agencia Nacional de Investigaci\'{o}n y Desarrollo (Chile), Ministerio de Ciencia, Tecnolog\'{i}a e Innovaci\'{o}n (Argentina), Minist\'{e}rio da Ci\^{e}ncia, Tecnologia, Inova\c{c}\~{o}es e Comunica\c{c}\~{o}es (Brazil), and Korea Astronomy and Space Science Institute (Republic of Korea).

\bibliography{refs.bib}
\end{document}